
\magnification=1200
\font\tit=cmssbx10 scaled \magstep2
\hsize=16.2truecm \vsize=23.5truecm
\nopagenumbers
\footline={\ifnum\pageno>0\hfil\rm\number\pageno\hfil\fi}

\def\Re{{\rm Re}} \def\Im{{\rm Im}}
\def\cldot{\!\cdot\!}

\def\*{\star}

\def\hf{{\textstyle {1\over 2}}}

\def\~#1{\mathaccent "7E #1}
\def\=#1{\mathaccent 22 #1}
\def\^#1{\mathaccent 94 #1}

\def\Q{{{\bf q}^2}}

\def\A{{\rm A}}

\def\gsim{\mathrel{\rlap{\lower0.5ex\hbox{$\sim$}}{\raise0.5ex\hbox{$>$}}}}
\def\refno#1{\item{[#1]}}
\def\sez#1{\bigbreak\vskip15truemm plus 5truemm minus 1truemm
 \leftline{\bf #1}\nobreak\bigskip}
\def\arrow#1{\hbox to #1truemm{\rightarrowfill}}
\def\limrel#1\for #2{\mathrel{\mathop{\kern0pt#1}\limits_{#2}}}
\def\fig#1 {\item{\bf Fig. #1.~}}
\pageno=0
\baselineskip=15pt
\null\vskip-1.5truecm\rightline{DFF - 165/7/92}
\vskip3mm\rightline{July 1992}
\vglue 2truecm
\centerline{\tit Planckian Vertices }
\vskip3mm\centerline{\tit on High Genus Riemann Surfaces}
\vglue 1.5truecm
\centerline{Alessandro Bellini}
\centerline{\it Dipartimento di Fisica dell'Universit\`a, Firenze
and INFN, Sezione di Firenze}
\centerline{Largo E. Fermi 2, I-50125 Firenze, Italy}
\vglue 3truecm
\centerline {\bf Abstract}
\bigskip
We suggest a method to compute leading contribution at Planckian energies
for superstring scattering amplitudes of any genus. In particular we test
the method at one-loop level by comparison with previous result for the
Regge trajectory renormalization. Modular invariance of these asymptotic
terms are also discussed.
\vfill\break
\sez{1. Introduction}

\baselineskip=18pt

Gravitational physics at Planckian energies has been studied in the past
through superstring scattering amplitude at fixed angle [1] or in the Regge
regime [2,3] and it is still under discussion [4,5] in order to understand
the role of short distances in string theory and quantum gravity.

Most of the interest in superstring originates in its unique role as a finite
[6] quantum theory containing gravity. Since gravitational interactions couple
to energy, the Regge regime of very large center-of-mass energies $\sqrt s$
and fixed transferred momentum $\sqrt -t$ turns out to be very promising.

It has been clarified [7] that in this asymptotic limit only pinched Riemann
surfaces give the main contribution to the Polyakov correlation function; in
a different language only particular regions of modular space contribute.
It was also shown that the Koba-Nilsen variables of external fast particles
get closer and closer as the energy is incresed, their distance being of the
order $1/s$.

In a previous paper [8] the relation between high energy and O.P.E. on the
string world sheet was exploited to introduce a reggeon emission vertex
$V_{\alpha,\~k}(z)$ suitable to study the Regge behavior of superstring
amplitudes.
The main achievement was the construction of a  slightly modified vertex
operator for inserting on the Riemann surface all the states lying on the
leading Regge trajectory.
Although the method could be used to compute Regge amplitudes of arbitrary
genus only tree amplitudes were considered explicitly [9].

Recently the one-loop correction to the graviton trajectory has been
reconsidered [10] due to its contribution to the superstring eikonal.
Starting from the one-loop superstring amplitude, after a proper subtraction
of the leading Regge cut term, the first subleading high-energy contribution
at large impact parameter has been obtained.

In this letter we present an alternative derivation of the same integral
representation which makes use of the Regge emission vertices.
Since these vertices describe external Regge poles in the angular momentum
plane, and the trajectory renormalization at first order arises as a double
pole correction, it is natural to compute this one by the insertion of two
Regge vertices on the torus world-sheet.
It turns out that the latter is a simple method to study the Regge behavior of
string amplitudes of any genus.
\bigskip
\sez {2. Off-shell vertex technique}

In a previous paper [10] we have investigated the pole and cut regions
of a four point amplitude directly on the torus, and we have found that in
the pole region the external punctures are close in pairs, because
$$ \vert \nu_a -\nu_d\vert\sim\vert \nu_b-\nu_c\vert\sim {1\over \sqrt {s}}.$$

This feature was used in ref. [9] as the starting point for the definition
of Regge vertices attached to the punctures.

We expect to obtain in a natural way the entire trajectory renormalization
at one loop by inserting two gravireggeon vertices on the torus world-sheet as
shown in fig.1a. This will turn out to be in agreement with the stationary
phase method applied to the four point amplitude at high energies used by
Sundborg in his analysis [11].

The gravireggeon emission vertex takes the form
$$ V_{\alpha ,\~k}
(z,\=z)=\int {d^2\delta\over\vert\delta\vert ^2}\vert\delta\vert ^\Q
V_R(z,\delta)V_L(\=z,\=\delta)\eqno (2.1)$$
where  $V_R$ is given in term of the right moving operators by
$$\eqalign{
V_R(z,\delta)&=:\left[-i\~k\cldot\partial_z X_R(z)+\~k\cldot\psi_R(z)
\^q\cldot\psi_R(z)\right] e^{\displaystyle i\^q\cldot X_R(z)}:\cr
&\phantom {=}\ \ \^q\equiv q+\~k\delta {\partial\over\partial z}\cr}
\eqno (2.2)$$
and similarly for $V_L$ with the obvious substitution.
We remind that $q$ is the off-shell momentum transfer coupled with to the
surface (fig 1a), while $\~k$ is the external momentum flowing in the fast
legs, which in Eq. (2.2) plays the role of polarization vector,
with $\~k\cldot q=0$.

We now describe how to compute the correlation functions of off-shell vertices
(2.1) over a genus one Riemann surface in superstring type II formalism.
First of all we remind that
on the torus there are 4 spin structures, one odd corresponding to
periodic-periodic boundary condition for all world-sheet spinors $\psi$ and
three even ones, containing at least one anti-periodic boundary condition.
These condition must be realized independently for left and right components.

The amplitude for two external punctures turns out to be an integral over
modular space of the following density
$$ \~A=\sum_{(a,\=a)}M_a \=M_{\=a}C_{a,\=a}\ll V_{\alpha,\~k_1}
(z_1,\=z_1)V_{\alpha,\~k_2}(z_2,\=z_2)\gg_{a,\=a}\eqno (2.3)$$
where the factor $M_a \=M_{\=a}C_{a,\=a}$ is the contribution of spin structure
$(a,\=a)$ to the one-loop partition function, that for even spin structures
[12] is:
$$M_a={\vartheta\left[ a\right]\left(0\vert\tau\right)^4\over
\vartheta_1^\prime\left(0\vert\tau\right)^4}\quad\quad\quad a=(a_1,a_2)=
\{ (0,0),(0,1),(1,0)\}$$
$$ \vartheta_{00}(\nu\vert\tau)=\vartheta_3(\nu\vert\tau)\quad\quad
\vartheta_{01}(\nu\vert\tau)=\vartheta_4(\nu\vert\tau)\quad\quad
\vartheta_{10}(\nu\vert\tau)=\vartheta_2(\nu\vert\tau)$$
and the factor $C_{a,\=a}$, discussed by Alvarez-Gaume and Vafa [13],
is for type II superstring only a phase $(-)^{a_1+a_2+\=a_1+\=a_2}$.
We can limit ourselves to even spin structures
because to get a non zero result from odd spin structures we need at least
six external particles. We recall [6] that partition function,
tadpole insertion, self-energy correction and 3-point amplitude vanishes at
one-loop level due to the Jacobi relation for theta functions [14]
$$\vartheta_3^4(0\vert\tau)=\vartheta_2^4(0\vert\tau)+
\vartheta_4^4(0\vert\tau)$$
and the constraint of on-shell vertices. We shall see that this result cannot
be extended to off-shell vertices.

We proceed in the computation of Eq.(2.3) by introducing the sum over loop
momenta to exhibit factorization of left and right modes which is allowed due
to the absence of non-holomorphic odd spin structures:
$$\ll V_{\alpha_1,\~k_1}(z_1,\=z_1)V_{\alpha_2,\~k_2}(z_2,\=z_2)\gg_{a,\=a}=
\int d^{10}p {d^2\delta_1 d^2\delta_2\over \vert\delta_1\delta_2\vert ^2}
\vert\delta_1\delta_2\vert ^\Q \Lambda_{\displaystyle q_1,q_2}(z_i,\delta_i)
\=\Lambda_{{\displaystyle q_1,q_2}}(\=z_i,\=\delta_i)$$
where we have introduced the shorthand notation
$\Lambda_{{\displaystyle q_1,q_2}}(z_i,\delta_i)$ to mean the following chiral
component
$$\eqalign {\Lambda_{{\displaystyle q_1,q_2}}^a&=\bigl\langle\bigl( -
{\partial\over\partial\delta_1}
+:\~k_1\cldot\psi (z_1)\^q_1\cldot\psi (z_1):\bigr)\bigl( -
{\partial\over\partial\delta_2}
+:\~k_2\cldot\psi (z_2)\^q_2\cldot\psi (z_2):\bigr)\bigr\rangle_a\cr
&\phantom {=}
\exp\left[ i\pi p^2\tau +2i\pi p\cldot\left(\^q_1 z_1+\^q_2 z_2\right)-
\^q_1\cldot\^q_2 G(z_1,z_2)\right]\cr}\eqno (2.4)$$
and the chiral $X$-propagator $ G(z_1,z_2) $ [12] or correlator
of chiral bosonic fields
$$ G(z_1,z_2)=\langle X(z_1)X(z_2)\rangle =-\ln E(z_1,z_2)$$
over non zero modes. It may be useful to recall that the prime form $E$ takes
a simple form for the torus
$$E(z_1,z_2)=
{\vartheta (z_1-z_2\vert\tau )\over\vartheta_1^\prime (0\vert\tau )}$$
and that the Dirac propagator for even spin structures
$$\langle \psi^\mu (z_1)\psi^\nu (z_2) \rangle_a=g^{\mu\nu}S_a(z_1,z_2)$$
is given by the Szeg\"o kernel
$$ S_a(z_1,z_2)=
{\vartheta [a] (z_1-z_2\vert\tau )\vartheta_1^\prime (0\vert\tau)\over
\vartheta [a] (0\vert\tau )\vartheta_1 (z_1-z_2\vert\tau)}.\eqno (2.5)$$
Keeping into account the normal ordering condition and the transversality of
$\~k_i$ with respect to $q_i$, the fermion correlator of equation (2.4)
gives for a specific spin structure the following result
$$\eqalign{\Lambda_{\displaystyle q_1,q_2}^a
&=\Bigl[{\partial^2\over\partial\delta_1\partial\delta_2}+
\~k_1\cldot\~k_2 q^2 S_a(z_1,z_2)^2 +(\~k_1\cldot\~k_2)^2\delta_1\delta_2
{\partial S_a(z_1,z_2)\over\partial z_1}
{\partial S_a(z_1,z_2)\over\partial z_2}+\cr
&\phantom {=}
-(\~k_1\cldot\~k_2)^2\delta_1\delta_2 S_a(z_1,z_2)
{\partial^2 S_a(z_1,z_2)\over\partial z_1\partial z_2}\Bigr]
\exp\Bigl[i\pi p^2\tau+2i\pi p\cldot q (z_1-z_2)+\cr
&\phantom {=}
+2i\pi p\cldot (\~k_1\delta_1+\~k_2\delta_2)+q_1\cldot q_2\ln E(z_1,z_2)+
\~k_1\cldot\~k_2\delta_1\delta_2{\partial^2\over\partial z_1\partial z_2}
\ln E(z_1,z_2)\Bigr]\cr}\eqno (2.6)$$
Summing now over the spin structures of the right and left sector we have
for the
\noindent density $\~A$
$$\~A=\int d^Dp\int\vert\delta_1\delta_2\vert^\Q\Bigl(\sum_{(a)}(-)^{a_1+a_2}
M_a\Lambda^a\Bigr)\bigl( c.c.\Bigr){d^2\delta_1\over \vert\delta_1\vert^2}
{d^2\delta_2\over \vert\delta_2\vert^2}\eqno (2.7)$$
and using the general Riemann identities we obtain a non-zero result only from
the last two terms in the first factor of Eq. (2.6), proportional to the
off-shell deformations $\delta_1 ,\delta_2$. This was to be expected because
going on-shell we should recover a null result for the two point amplitude.
We now discuss in more detail this last contribution
$$-(\~k_1\cldot\~k_2)^2\delta_1\delta_2\sum_{(a)}(-)^{a_1+a_2}
S_a^2(z_1-z_2){\partial^2\over\partial z_1^2}\ln S_a(z_1-z_2)$$
which in terms of theta function reads
$$-(\~k_1\cldot\~k_2)^2\delta_1\delta_2\sum_{(a)}{(-)^{a_1+a_2}
\vartheta [a](0\vert\tau )^2\over\vartheta_1^\prime (0\vert\tau )^2
\vartheta_1(z_1-z_2)^2}\Bigl(\vartheta^\prime [a](z_1-z_2\vert\tau )^2-
\vartheta [a](z_1-z_2\vert\tau )\vartheta^{\prime\prime}[a](z_1-z_2\vert\tau )
\Bigr)\eqno (2.8)$$
Using the identity
$$\sum_{(a)}(-)^{a_1+a_2}\vartheta [a](0\vert\tau)^2\vartheta [a](u\vert\tau)
\vartheta [a](v\vert\tau)=2\vartheta_1^2({u+v\over 2}\vert\tau)
\vartheta_1^2({u-v\over 2}\vert\tau)$$
and its derivatives computed for $u=v$, the above expression simplifies to
$ -2(\~k_1\cldot\~k_2)^2\delta_1\delta_2$ and similarly for the left sector.
Finally integrating Eq. (2.7) over the ten dimensional loop momenta and
introducing the toroidal compactification factor $F_c$ we obtain the
following amplitude
$$\eqalign{
A&=4g^2(\~k_1\cldot\~k_2)^4\int {d^2\tau\over\tau_2^5}F_c(\tau,R_c)d^2z
d^2\delta_1 d^2\delta_2\vert\delta_1\delta_2\vert^\Q
e^{\displaystyle
{-2\~k_1\cldot\~k_2{2\pi\over\tau_2}\Im\delta_1\Im\delta_2}}\cr
&\phantom {=g^2}
\Biggl\{\biggl\vert
{\vartheta_1(z\vert\tau)\over\vartheta_1^\prime (0\vert\tau)}\biggr\vert
e^{\displaystyle {-{\pi\over\tau_2}\Im^2 z}}\Biggr\}^{-2\Q}
e^{\displaystyle {+2\~k_1\cldot\~k_2 \Re\Big[
\delta_1\delta_2{\partial^2\ln\vartheta_1(z\vert\tau)\over\partial z^2}
\Big]}}\cr}\eqno (2.9)$$
We note the agreement of the above expression with Eq. (5.6) of ref. [11],
where $\delta_1 ,\delta_2$ are the variables of the saddle-point approximation
used by Sundborg to compute the double-pole correction (in the angular
momentum plane) starting from the four dilatons scattering amplitude on
the torus. It is worthwhile to discuss more the last expression.

Generally a superstring amplitude is Regge behaved and admits the following
expansion
$$\A=\sum_{i,k} s^{\alpha_{i,k}(t)}(\ln s)^{k}\beta_{i,k}(t)$$
where, in formal analogy with O. P. E. in quantum field theory,
the physical quantities are the Regge trajectories $\alpha_{i,k}(t)$
(anomalous dimension) and their residues $\beta_{i,k}(t)$ (non trivial
vacuum expectation values) and the large energy replace the short distance
expansion.

Indeed for the amplitude (2.9) the polarization-like vectors $\tilde k$
give rise to the large energy factor $s=2\tilde k_1\cldot\tilde k_2$ in the
exponent and require the smallness of $\delta_1\delta_2$.
The latter are interpreted like distances between the Koba-Nilsen variables
of the external fast legs (see fig. 1a) $\delta_1=\nu_a-\nu_d$ and
$\delta_2=\nu_b-\nu_c$, where we use the same notation of ref. [10] for
comparison of the result.

The aforementioned Regge behavior of $\A$ is naively \footnote {*}{the actual
asymptotic behavior is the one typical of Regge cut $s^{3+\hf t}$ due to
an overlap of the relevant integration regions, and make expression (2.9)
ill-defined without a proper subtraction as discussed in ref. [10].}
factorized-out integrating
$\delta_1$ and $\delta_2$ over two small disks of radius $\sim 1/\sqrt s$
with the result
$$\A=g^2\beta_{1}(t) s^{2+t}\ln s  + {\rm subleading\  terms}$$
where the logaritmic factor is typical of the rapidity integral in
the Regge-Gribov calculus.

Adding this term to the asymptotic tree amplitude $s^{2+t}\beta_{0}(t)$ and
assuming exponentiation at higher genus for the full amplitude
$$\A_f= s^{2+t+g^2{\beta_1(t)\over \beta_0(t)}+...}\beta_0(t)+
{ \rm higher\ order\ terms},$$
we identify the first order renormalization to the graviton trajectory.

Finally we show that the integral  representation (2.9) is modular
invariant. We refer to [6,12] for the definition of a modular transformation,
we only quote here that the quantity in braces in (2.9) is a modular form
$\chi \left( z\vert\tau\right)$ of weight one [14], with the property
$$ \chi\left(z'\vert \tau '\right)=
\vert c\tau + d\vert^{-1}\chi\left(z\vert\tau\right)$$
while the theta function transform as
$$\vartheta_1\left(z'\vert\tau '\right)=
\eta {\left( c\tau +d\right)}^\hf \exp {\left(i\pi c z^2\over c\tau +d\right)}
\vartheta\left(z\vert\tau\right),$$
where $\eta$ is a constant phase.

The exponent in (2.9), proportional to the large kinematical variable $s$,
is transformed into
$$-{2\pi {\vert c\tau +d\vert}^2\over\tau_2}\Im {\delta_1\over c\tau +d}
\Im {\delta_2\over c\tau +d} +\Re\Big[\delta_1\delta_2 {2i\pi\over c\tau +d}
+\delta_1\delta_2{\partial^2\ln\vartheta_1(z\vert\tau)\over\partial z^2}
\Big]$$
and it is matter of simple algebra to show the invariance of the integrand.
The only change is in the radius of the small disks's integration regions.
We get finally the modular invariance of $\beta (t)$ noticing that the actual
extension of these circular regions is irrelevant [10] for the leading
asymptotic behavior.

We conclude stating that the off-shell vertices used before are a useful
concept in high energy superstring computations, when only pinched Riemann
surfaces are relevant for the asymptotic behavior.

We finally point out that the contribution (2.9), which corresponds to the
pinched surface (fig. 1b), is the same as the one obtained by
the Regge-Gribov diagram with two external Regge poles shown in fig. 1a.
This is in general very different from the Feynman diagram calculation.
In particular the $\ln s$ factor, typical of the above double-pole
correction, is produced (in a covariant gauge) only as a subleading term of a
box diagram [10] and not by the insertion of a renormalized graviton
propagator as fig. 1b could erroneusly suggest.
\vglue 4truecm
\noindent {\bf Aknowledgments}

\noindent The author would like to express his gratitude to M.Ademollo and
M.Ciafaloni for enlightening discussions.
\vfill
\eject
\sez {References}
\vskip0.5cm
\refno {1} D.J.Gross and P.F.Mende, Phys. Lett B 197 (1987) 129;
Nucl. Phys. B303 (1988) 407.
\refno {2} D. Amati, M. Ciafaloni and G. Veneziano, Phys. Lett. B197
(1987) 81; Int. J. Mod. Phys. 3A (1988) 1615.
\refno {3} D. Amati, M. Ciafaloni and G. Veneziano, Nucl. Phys. B347
(1990) 550.
\refno {4}D. Amati, M. Ciafaloni and G. Veneziano, {\it Planckian scattering
beyond the semiclassical approximation}, CERN preprint TH.6395/92.
\refno {5} E. Verlinde and H. Verlinde, {\it Scattering at planckian energies},
Princeton preprint PUTP-1279 (1991); see also R.Kallosh, {\it Geometry of
scattering at Planckian energies}, Stanford preprint SU-ITP 903 (1991).
\refno {6} M. B. Green, J. H. Schwarz and E. Witten,
{\it Superstring theory} (Cambridge University Press, Cambridge, 1987).
\refno {7} E.Gava, R.Iengo and C.J.Zhu, Nucl. Phys. B323 (1989) 585;
A. Bellini, G. Cristofano, M. Fabbrichesi and K. Roland,
Nucl. Phys. B356 (1991) 69.
\refno {8} M. Ademollo, A. Bellini and M. Ciafaloni, Phys. Lett.
B223 (1989) 318.
\refno {9} M. Ademollo, A. Bellini and M. Ciafaloni, Nucl. Phys.
B338 (1990) 114.
\refno {10} M. Ademollo, A. Bellini and M. Ciafaloni, {\it Superstring
One-Loop and Gravitino Contribution to Planckian Scattering}, Nordita
preprint 92/45 (July 1992).
\refno {11} B. Sundborg, Nucl. Phys. B306 (1988) 545.
\refno {12} E. D`Hoker and D. H. Phong, Rev. Mod. Phys. 60 (1988) 917.
\refno{13}  L. Alvarez-Gaume, G. Moore and C. Vafa,  Commun. Math.
Phys. 106 (1986) 1.
\refno{14} D. Mumford {\it Tata Lectures on Theta Functions}, (2 vol.)
Birkhauser (1983).
\vfill\eject
\bye